\documentclass[Times,8pt,aps,prl,twocolumn,showpacs,amsmath,amssymb,floatfix,footinbib,superscriptaddress]{revtex4-1}
\usepackage{amsmath,natbib,graphicx,subfigure,amssymb,graphics,amsmath,mathrsfs,CJK,color}
\usepackage{multirow,fancyhdr,color,bm,tabularx,psfrag,geometry,dcolumn,datetime}
\usepackage[colorlinks=true,linkcolor=blue,urlcolor=blue,citecolor=blue]{hyperref}
\usepackage[mathlines]{lineno}
\def\footnoterule{\kern -1mm \hrule width 5.8cm \kern 2.2mm}%
\usepackage{tikz,xcolor,hyperref}
\definecolor{lime}{HTML}{A6CE39}
\DeclareRobustCommand{\orcidicon}{%
 \begin{tikzpicture}
 \draw[lime, fill=lime] (0,0)
    circle [radius=0.16]
    node[white] {{\fontfamily{qag}\selectfont \tiny ID}};\draw[white, fill=white] (-0.0625,0.095)
    circle [radius=0.007];
 \end{tikzpicture}
\hspace{-2mm}}
\foreach \x in {A, ..., Z}
{\expandafter\xdef\csname orcid\x\endcsname{\noexpand\href{https://orcid.org/\csname orcidauthor\x\endcsname}{\noexpand\orcidicon}}}

\begin{document}


\title{Effect of Spontaneously Generated Coherence and detuning on 2D Atom Localization in Two Orthogonal Standing-wave Fields}

\thanks{Supported by the National Natural Science Foundation of China ( Grant Nos. 61205205 and 6156508508 ),
the General Program of Yunnan Provincial Research Foundation of Basic Research for application, China ( Grant No. 2016FB009 )
and the Foundation for Personnel training projects of Yunnan Province, China ( Grant No. KKSY201207068 ).}


\author{Shun-Cai Zhao\orcidA{}}
\email[Corresponding author: ]{ zhaosc@kmust.edu.cn }
\affiliation{Department of Physics, Faculty of Science, Kunming University of Science and Technology, Kunming, 650500, PR China}

\author{Qi-Xuan Wu}
\affiliation{College English department, Kunming University of Science and Technology, Kunming, 650500}

\author{Ai-Ling Gong}
\affiliation{Department of Physics, Faculty of Science, Kunming University of Science and Technology, Kunming, 650500, PR China}


\begin{abstract}
Two-dimensional(2D)atom localization via the spontaneously generated coherence (SGC) and detunings associated with the probe and standing-wave driving fields in a
three-level V-type atomic system is investigated. In the gain process, two equal and tunable peak maximums of position distribution in the x-y plane via the detunings are observed.
However, one decreasing and other increasing peak maximums in the absorption process via the spontaneously generated coherence(SGC) are achieved in the quadrants I and III of the x-y plane. A better
resolution and more novelty for 2D atom localization in our scheme is obtained.
\end{abstract}



\maketitle
Subwavelength atom localization has become an active research topic from both the theoretical and experimental points of view, because
of its applications in the laser cooling and trapping of neutral atoms$^{[1,2]}$, atom lithography$^{[3]}$, measurement of atomic wave function$^{[4]}$, etc.. Several schemes utilized a standing-wave
field$^{[5]-[14]}$ for atom localization have been proposed since the position-dependent atom-field interaction, such as the effect of
spontaneous emission$^{[6,7]}$, population$^{[8,10,12,14]}$, absorption$^{[9]}$, the entanglement between the atom`s internal
states and its position$^{[15]}$, phase shift of the field$^{[16]}$ and gain$^{[17]}$ on the atomic position information was discussed. Recently, one-dimensional (1D) sub-half-wavelength atom localization was achieved$^{[17-25]}$ by interacting with two standing-wave fields. The resulting quantities, for instance, the homodyne detection$^{[5]}$, Raman gain process$^{1[7]}$, phase shift$^{[1,18]}$, probe field absorption$^{[9,20]}$,
quantum trajectories$^{[21]}$, upper level population$^{[8,21]}$, dual quadrature field$^{[22]}$, two-photon spontaneous emission$^{[23]}$, coherent population
trapping$^{[24]}$ and the superposition of two standing-wave fields$^{[25]}$ can provide the one-dimensional(1D) atomic position information. In the same way, the 2D atom localization was achieved utilizing multiple simultaneous quadrature measurement, the population in the upper state or in any ground state measurement, probe absorption measurement and incorporating the quantum interference phenomenon measurement, respectively$^{[26-29]}$. And other schemes$^{[30-34]}$ for 2D atom localization were also proposed in recent years. Such as in the coherently driven cycle-configuration and a five-level M-type atomic systems are proposed for 2D atom localization based
on controllable spontaneous emission$^{[30,32]}$.

However, spontaneously generated coherence (SGC) refers to the
interference of spontaneous emission channels$^{[35]}$ firstly
showed by Agarwal$^{[36]}$ in a degenerate $\Lambda$-type
three-level atom system, and which plays an important
role$^{[37-45]}$ in lasing without population inversion,
coherent population trapping(CPT), group velocity reduction, ultra
fast all-optical switching and transparent high-index materials,
high-precision spectroscopy and magnetometer and modified quantum
beats, etc. Inspired by these investigations, we here utilize the
two interference of spontaneous emission channels to explore an
interesting scheme for 2D atom localization. When the two orthogonal
standing-wave fields drive the same atomic transition in the
$\Lambda$-type atom, two equal and tunable peaks of the position
probability distribution occur in quadrants I and III when
manipulating the detunings corresponding to the probe and the
coupled orthogonal standing-wave fields. However, when the
SGC intensities were varied, one decreasing and other increasing peak maximums
of 2D atom localization in the quadrants I and III of
x-y plane appear. Comparing our atomic scheme and the traditional 2D
atom localization schemes $^{[26-28]}$, the results show noticeably
an increasing spatial resolution and high precision,
and manifest much more flexibility and novelty.

We consider a V-type system[see Fig.1] which has a lower state
$|a\rangle$ and two excited states $|c\rangle$ and $|b\rangle$.
A weak probe field with frequency
$\nu_{p}$ and Rabi frequency $E_{p}$ of
$\Omega_{p}=E_{p}\mu_{ab}/2\hbar$ drives the
transition $|a\rangle$$\leftrightarrow$$|b\rangle$ with frequency
$\omega_{ab}$. And a strong coherent coupling
standing-wave field $E_{x,y}$ with the same frequency $\nu_{c}$ is
simultaneously applied to the transition
$|a\rangle$$\leftrightarrow$$|c\rangle$ with frequency
$\omega_{ac}$. The Rabi frequency $\Omega_{c}(x,y)$ corresponding to
the composition of two orthogonal standing waves is
 $\Omega_{c}(x,y)=E_{x,y}\mu_{ac}/2\hbar$=$\Omega_{0}[sin(\kappa_{1}x+\delta)+sin(\kappa_{2}y+\eta)]$.
Where $\mu_{ab}$ and $\mu_{ac}$ are the corresponding dipole matrix
elements, $\kappa_{i}=2\pi/\lambda_{i},(i=1,2)$ is the wave vector
with wavelengths $\lambda_{i},(i=1,2)$ of the corresponding standing
wave fields. For simplicity, we assume
$\Omega_{p}$ and $\Omega_{0}$ to be real. $\Delta_{p}$=$\omega_{ab}-\nu_{p}$, and
$\Delta_{c}$=$\omega_{ac}-\nu_{c}$ are the detunings of the two
corresponding fields, respectively. $2\gamma_{1}$ and
$2\gamma_{2}$ are the spontaneous decay rates of the excited state
$|c\rangle$ and $|b\rangle$ to the ground states $|a\rangle$,
respectively. When the two excited levels $|c\rangle$ and
$|b\rangle$ are closely spaced such that the two transitions to the
lower state interact with the same vacuum mode, SGC can be present.
\vskip 0.5\baselineskip
\vskip 4mm
\centerline{\includegraphics{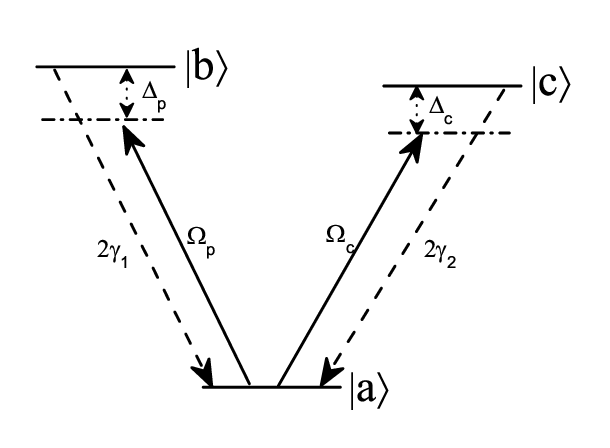}}
\vskip 2mm
\centerline{\footnotesize \begin{tabular}{p{6.5 cm}}\bf Fig.\,1. \rm
Level diagram of the atomic system: Two orthogonal standing waves fields having position-dependent Rabi $\Omega_{c}(x,y)$
corresponding to the atomic transition from $|c\rangle$ to
$|a\rangle$. A weak probe field $\Omega_{p}$ couples the transition from $|a\rangle$ to $|b\rangle$. $2\gamma_{1}$ and
$2\gamma_{2}$ are the atomic decay rates.
\end{tabular}}
Even though there is a principle difficulty in the realization of atomic
interference via spontaneous emission for atomic transitions in free space,
the technique of photonic band-gap materials and semiconductor quantum
dots was proposed to realize the interference$^{[46]}$.
Since the pseudophotonic band-gap structures have a special purpose$^{[47]}$,
which can allow the propagation of electromagnetic waves with a certain polarization
and strongly forbid those with orthogonal polarizations.
So, the spontaneous decay with forbidden polarization
is also forbidden and this allows us to remove the cancellation
of different contributions originated from different polarizations
of spontaneous emission occurring in free space, which is just what we need.

Along the directions of the standing waves, the center-of-mass position
of the atomic is assumed to be nearly constant. Hence, the Raman-Nath
approximation is applied and the kinetic energy of
the atom in the Hamiltonian is ignored$^{[27,48]}$. Under the electric dipole and
rotating-wave approximation, we write the systematic density matrix in the
interaction picture involving the SGC as$^{[49-50]}$

\begin{eqnarray}
&\dot{\rho_{cc}}=&-2\gamma_{1}\rho_{cc}+i\Omega_{c}(\rho_{ac}-\rho_{ca})-p\sqrt{\gamma_{1}\gamma_{2}}(\rho_{cb}+\rho_{bc}), \\
&\dot{\rho_{bb}}=&-2\gamma_{1}\rho_{bb}+i\Omega_{p}(\rho_{ab}-\rho_{ba})-p\sqrt{\gamma_{1}\gamma_{2}}(\rho_{cb}+\rho_{bc}),  \\
&\dot{\rho_{ac}}=&-(\gamma_{1}+i\Delta_{c})\rho_{ac}-p\sqrt{\gamma_{1}\gamma_{2}}\rho_{ab}+i\Omega_{c}(\rho_{cc}-\rho_{aa})\nonumber\\&&+i\Omega_{p}\rho_{bc},  \\
&\dot{\rho_{cb}}=&i\Omega_{c}\rho_{ab}-i\Omega_{p}\rho_{ca}-i(\Delta_{p}-\Delta_{c})\rho_{cb}-(\gamma_{1}+\gamma_{2})\rho_{cb}\nonumber\\&&-p\sqrt{\gamma_{1}\gamma_{2}}(\rho_{cc}+\rho_{bb}), \\
&\dot{\rho_{ab}}=&-p\sqrt{\gamma_{1}\gamma_{2}}\rho_{ac}-(\gamma_{2}+i\Delta_{p})\rho_{ab}+i\Omega_{c}\rho_{cb}\nonumber\\&&+i\Omega_{p}(\rho_{bb}-\rho_{aa}). 
\end{eqnarray}

along with the the requirement of closing system, i.e.,
$\rho_{aa}+\rho_{bb}+\rho_{cc}=1$ and $\rho_{ij}^{\ast}=\rho_{ji}$.
And in which the parameter p denoting the alignment of the two dipole
moments is defined as $p=\vec{\mu_{ab}}\cdot\vec{\mu_{ac}}/|\vec{\mu_{ab}}\cdot\vec{\mu_{ac}}|=cos\theta$
with $\theta$ being the angle between the two dipole moments
$\mu_{ac}$ and $\mu_{ab}$, which can be a random angle between 0 and
2$\pi$ except for 0 and $\pi$, and which is very sensitive for the
existence of the SGC effect. The terms with $p\sqrt{\gamma_{1}\gamma_{2}}$ represent the quantum interference
resulting from the cross coupling between spontaneous emission paths
$|a\rangle$-$|c\rangle$ and $|a\rangle$-$|b\rangle$. It should be
noted that only for small energy spacing between the two excited
levels are the interference terms in the systematic density matrix
significant; otherwise the oscillatory terms will average out to
zero and thereby the SGC effect vanishes.

\par The information about the 2D atomic position from the susceptibility$^{[22]}$ of the system at the probe field frequency
is what we want to get. The nonlinear Raman susceptibility $\chi$ is then given by

\begin{equation}
\chi=\frac{2N|\mu_{ab}|^{2}}{\epsilon_{0}\Omega_{P}\hbar}\rho_{ab}, \label{2}
\end{equation}

where N is the atom number density in the medium and $\mu_{ab}$ is the magnitude of the dipole-matrix element between $|a\rangle$ and
$|b\rangle$. $\epsilon_{0}$ is the permittivity in free space. The general steady-state analytical solution for $\rho_{ab}$ can be
written as

\begin{align}
\rho^{1}_{ab}=&\frac{-4B_{7}+\Omega_{p}\{-B_{1}\rho^{0}_{ac}+\Omega_{c}[-i(B_{0}-B_{5})p\rho^{0}_{ba} }{4[-p^{2}B_{8}+B_{10}p^{4}-4p^{6}+B_{9}(1+\Delta_{c}^{2}+2\Omega_{c}^{2})]}+ \label{3}
\end{align}

\noindent with
\begin{eqnarray}
&\rho^{0}_{ac}=&\frac{(\Delta_{p}-i)(\Delta_{p}-2i-\Delta_{c})\Omega_{c}-\Omega_{c}^{3}}{(2i+\Delta_{c}-\Delta_{p})[p^{2}+(\Delta_{p}-i)(\Delta_{c}-i)]+(\Delta_{c}-i)\Omega_{c}^{2}}, \\
&\rho^{0}_{ab}=&\frac{p\Omega_{c}(2i+\Delta_{c}-\Delta_{p})}{(2i+\Delta_{c}-\Delta_{p})[ip^{2}-i+ \Delta_{c}(1+i\Delta_{p})+\Delta_{p}]+(1+i\Delta_{c})\Omega_{c}^{2}}, \\
&\rho^{0}_{cb}=&\frac{ip\Omega_{c}^{2}}{(2i+\Delta_{c}-\Delta_{p})[p^{2}+(\Delta_{p}-i)(\Delta_{c}-i)]+(\Delta_{c}-i)\Omega_{c}^{2}}, \\
&\dot{\rho_{ab}}=&-p\sqrt{\gamma_{1}\gamma_{2}}\rho_{ac}-(\gamma_{2}+i\Delta_{p})\rho_{ab}+i\Omega_{c}\rho_{cb}+i\Omega_{p}(\rho_{bb}-\rho_{aa}).
\end{eqnarray}

where the parameters $B_{i},(i=0,\ldots,10)$ in Eq.(3) are given in
the Appendix, and we have set $\gamma_{1}=\gamma_{2}=\gamma$. All the parameters
are reduced to dimensionless units by scaling with $\gamma$.
Using Eq.(2), which consists of both real and imaginary
parts, i.e., $\chi=\chi^{'} + i\chi^{''}$. The imaginary part of the
susceptibility gives the absorption profile of the probe field which
can be written as
\begin{align}
\chi^{''}=\frac{2N|\mu_{ab}|^{2}}{\epsilon_{0}\hbar}Im[\frac{\rho_{ab}}{\Omega_{P}}]=\alpha
Im[\frac{\rho_{ab}}{\Omega_{P}}], \label{7}
\end{align}
where $\alpha=\frac{2N|\mu_{ab}|^{2}}{\epsilon_{0}\hbar}$. Here we
are interested in the 2D position measurement of the atom using the
absorption process of the probe field. The Eq.(7) is the main result
and reflects the conditional position probability distribution of
the atom$^{[22]}$. It can be seen that the probe absorption depends
on the position dependent the SGC intensities p and the detunings
associated with the probe and standing-wave driving fields,
therefore, we can obtain the 2D atomic position information by
manipulating the corresponding parameters.

In this section, we analyze the imaginary part $\chi^{''}$ of the
susceptibility depicting the absorption profile of the probe field
which directly reflects the 2D atomic conditional position probability
distribution, and then demonstrate a high resolution of 2D
atom localization via manipulation of the values of detunings or the
the intensities p of SGC. It is evident that $\chi^{''}$ depends on
the parameters of the field detunings and the intensities p of SGC,
which can be seen from Eq.(3) to Eq.(6). However, the analytical
expressions from Eq.(3) to Eq.(6)corresponding to the field detunings
and the intensities p of SGC is rather cumbersome.
Hence, the numerical approach was preferred to
analyze the the 2D atomic position probability distribution via $\chi^{''}$.
In the following discussion, the probe field absorption $\chi^{''}$
reflecting different position probability
distribution of the atom within optical wavelength domain is plotted.

Initially, the position dependent the probe field detuning
$\Delta_{p}$ is considered. We set $\Delta_{c}$=0, and the Rabi frequency,
$\Omega_{c0}$$=$$10\gamma$, $\Omega_{p0}$$=$$0.01\gamma$. Then a three-dimensional
plot depicting the effect of position-dependent the probe detuning
$\Delta_{p}$ for the imaginary susceptibility $\chi^{''}$ as a function of (x,y)
is shown in Fig.2.

\vskip 0.5\baselineskip
\vskip 4mm
\centerline{\includegraphics{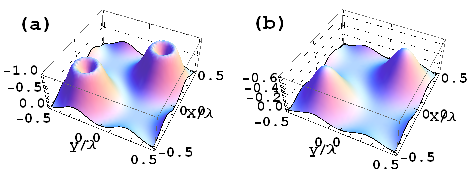}}
\centerline{\includegraphics{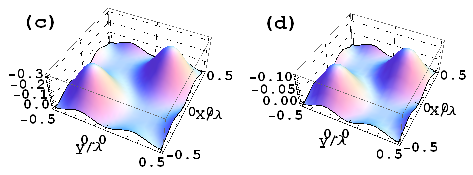}}
\vskip 2mm
\centerline{\footnotesize \begin{tabular}{p{7.5 cm}}\bf Fig.\,2. \rm
(Color online) Plots for 2D atom localization: $\chi^{''}$ as a function of
(x,y) in dependence on the probe detuning $\Delta_{p}$. (a)$\Delta_{p}$=21$\gamma$,(b)$\Delta_{p}$=30$\gamma$, (c)$\Delta_{p}$=40$\gamma$, (d)$\Delta_{p}$=60$\gamma$. Parameters are $\kappa_{1}$=$\kappa_{2}$=$2\pi$/$\lambda$, $\Omega_{c0}$=10$\gamma$, $\Omega_{p0}$=0.01$\gamma$, $\Delta_{c}$=0, $\theta$=0.5$\pi$, and where $\gamma$ is the scaling parameter.
\end{tabular}}

As shown in Fig.2(a), a craterlike pattern occurs in
the quadrants I and III and which leads to the localization of the atom at
these circles with peak maxima being -1. An equal maximum probability
pattern is gotten in the standing-wave plane. When $\Delta_{p}$=30$\gamma$
in Fig.2(b), the probability distribution changes into two spikelike patterns in
the quadrants I and III of x-y plane. And the peak maxima declines sharply to -0.6.
Increasing the probe detuning $\Delta_{p}$ repeatedly to 30$\gamma$,40$\gamma$ in Fig.2(c) and (d),
the spikelike pattern remains but its peak maxima continues to decrease. And the corresponding
maximum probabilities of 2D atom localization are -0.3, -0.1 in Fig. 2(c)and (d), respectively.
So, when the probe field off-resonantly couples $|a\rangle$$\leftrightarrow$$|b\rangle$ sharply, the
equal maximum probability of 2D atom localization decreases, and the resolution reduced.

\vskip 4mm
\centerline{\includegraphics{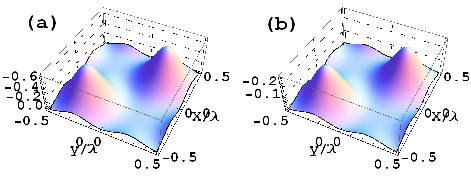}}
\centerline{\includegraphics{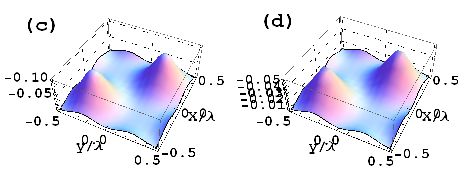}}
\vskip 2mm
\centerline{\footnotesize \begin{tabular}{p{7.5 cm}}\bf Fig.\,3. \rm
(Color online) Plots for 2D atom localization: $\chi^{''}$ as a function of
(x,y) in dependence on the probe detuning $\Delta_{c}$. (a)$\Delta_{c}$=$-1\gamma$,(b)$\Delta_{c}$=$-5\gamma$, (c)$\Delta_{c}$=$-10\gamma$, (d)$\Delta_{c}$=$-15\gamma$. $\Delta_{p}$=30$\gamma$. Other parameters are the same as in Fig.2.
\end{tabular}}

The coupled standing-wave field detuning $\Delta_{c}$ is another considered role in 2D atom localization.
In this case, we set $\Delta_{p}$=30$\gamma$ and the behavior of the 2D
localization is modulated by the coupled standing-wave field detuning
$\Delta_{c}$. In Fig.3, the coupled standing-wave field detuning is turned to
-1$\gamma$, -5$\gamma$, -10$\gamma$ and -15$\gamma$, respectively. And all the other
parameters are the same as those in Fig.2. The spikelike pattern occurs in
the quadrants I and III of the standing-wave plane from
Fig.3(a) to 3(d). Compared to Fig.2, we note that spikelike becomes much more acuity and
with the alike decreasing peak maxima pattern from Fig.3(a) to 3(d). It's worth noting that we obtained
the position distribution in the gain process.

As the above observed, we got a clear dependence of the detunings corresponding to the
probe and the coupled standing wave fields on the 2D atom
localization. Finally, we consider the position distribution dependent the
intensities p of SGC, and the corresponding intensities of SGC
p=$cos\theta$ is set as (a)$\theta=\pi/2.1$, (b)$\theta=\pi/2.3$,
(c)$\theta=\pi/2.5$, (d)$\theta=\pi/2.7$, as shown in Fig.4 from (a)
to (d), respectively. And their corresponding contour plots are shown
in Fig.5. Here we set the detunings corresponding to the probe and the coupled standing wave fields are
$\Delta_{P}$=30$\lambda$, $\Delta_{c}$=15$\lambda$, and set $\Omega_{p0}$=0.1$\gamma$. All the other
parameters used here are the same as in Fig.2.

\vskip 4mm
\centerline{\includegraphics{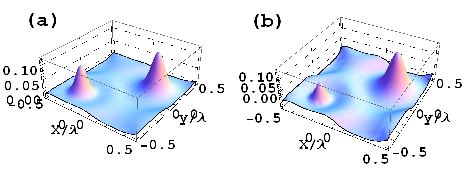}}
\centerline{\includegraphics{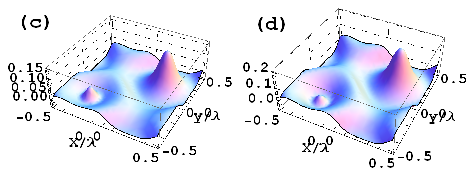}}
\vskip 2mm
\centerline{\footnotesize \begin{tabular}{p{7.5 cm}}\bf Fig.\,4. \rm
(Color online) Plots for 2D atom localization: $\chi^{''}$ as a function of
(x,y) in dependence on the intensities p of SGC. (a)$\theta$=$\pi/2.1$,(b)$\theta$=$\pi/2.3$, (c)$\theta$=$\pi/2.5$, (d)$\theta$=$\pi/2.7$. $\delta$=0, $\eta$=$\pi/12$, $\Delta_{c}$=15$\lambda$, $\Delta_{P}$=30$\lambda$, $\Omega_{p0}$=0.1$\gamma$. All the other parameters are the same as in Fig.2.
\end{tabular}}

It is interesting to compare the results shown in Figs.4
and those shown in Figs.2, in Figs.3. In Fig.4, we obtain two potential positions for the
atom localization with one having a higher probability than
the other in the absorption process. The peak maxima of the imaginary susceptibility
are no longer equal in the four quadrants. As can be seen from Fig.4(a), the peak
maxima in quadrant I is higher than the other in quadrant III. When the intensities of SGC
p=$cos\theta$ with $\theta$ be set as $\theta=\pi/2.3$ in Fig.4(b), the hight different
enhances because of the decrease of the peak maxima in quadrant III. The hight different aggravated in
Fig.4(c) and Fig.4(d). However, the reason for the aggravation is the same, which the peak maxima
is increasing in quadrant I in Fig.4(c) and Fig.4(d) .
\vskip 4mm
\centerline{\includegraphics{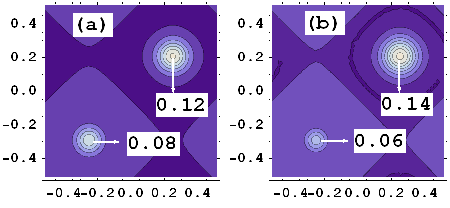}}
\centerline{\includegraphics{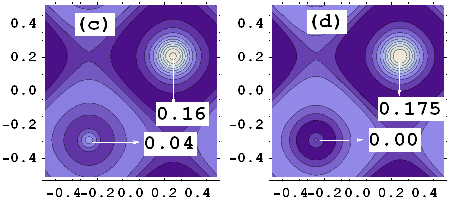}}
\vskip 2mm
\centerline{\footnotesize \begin{tabular}{p{7.5 cm}}\bf Fig.\,5. \rm
(Color online) Contour plots for 2D atom localization: $\chi^{''}$ as a function of
(x,y) in dependence on the intensities p of SGC. All the parameters in
(a) to (d) are the same as in Figs.4, respectively.
\end{tabular}}
In order to get a deeper sight of the role of the intensities p of SGC in 2D atom localization, we show their contour plots in
Fig.5. And the corresponding peak maxima values are labeled in Fig.5. From the contour plots shown in Fig.5(a) to Fig.5(d), the values
of peak maxima in quadrant I are 0.12, 0.14, 0.16 and 0.175, respectively. And the corresponding values of the peak maxima in quadrant
III are 0.08, 0.06, 0.04 and 0. The increasing values in quadrant I and decreasing values in quadrant
III demonstrate the effect of the intensities p of SGC on the 2D atom localization. An increasing resolution of 2D atom localization manipulated by the
intensities p of SGC is achieved.

In summary, a scheme of 2D atom localization based on the numerical
calculations and analyses is obtained. Because of the
spatial-dependent atom-field interaction, 2D atom localization can
be realized via measuring the probe absorption. We
investigated the conditional position probability distribution of an
V-type atom interacting with a weak probe field and two orthogonal
standing wave fields. By monitoring position dependent the detunings
corresponding to the probe and the standing-wave fields,
an equal and tunable peak maxima of position
probability distribution is obtained. While a better resolution of
position probability distribution is obtained via the intensity p of SGC.
The patterns of spikelike of imaginary susceptibility $\chi^{''}$ in
quadrant I and III for 2D atom localization vary differently.
And the peak maxima in quadrant I increases gradually while the other one in
quadrant III decreases. Compared with the mentioned previous schemes,
one peak maxima increasing and other one decreasing
for 2D atom localization are the novelty of our scheme.

\section*{Appendix}

The parameters $B_{i}$ in Eq.(3) are given by the following:

\begin{widetext}\begin{eqnarray}
&B_{0}=&P^{4}(8i+\Delta_{c}-3\Delta_{p})+\Omega_{c}^{2}(5i+\Delta_{c}-2\Delta_{p})[\Omega_{c}^{2}+(i+ \Delta_{p})(-2i+\Delta_{c}-\Delta_{p})]+ P^{2}[(i+\Delta_{p})(i+\Delta_{c})(8i+ \nonumber\\&&
        \Delta_{c}-3\Delta_{p})+2\Omega_{c}^{2}(5i+\Delta_{c}-4\Delta_{p})], \nonumber\\
& B_{1}=&2iP^{2}\{(8i+\Delta_{c}-3\Delta_{p})[P^{2}+(i+\Delta_{p})(i+\Delta_{c})]+\Omega_{c}^{2} (4i+\Delta_{c}-3\Delta_{p})\}, \nonumber\\
&B_{2}=&2P^{4}-[\Omega_{c}^{2}+(i+\Delta_{p})(-2i+\Delta_{c}-\Delta_{p})][\Omega_{c}^{2}+(i+\Delta_{c})(-3i+\Delta_{p})] \nonumber\\&&+ P^{2}[4+\Omega_{c}^{2}+\Delta_{p}(i+\Delta_{p})+\Delta_{c}(5i+\Delta_{p})], \nonumber\\
&B_{3}=&-iP^{4}(-4i+\Delta_{c}-3\Delta_{p})+2[\Omega_{c}^{2}+(i+\Delta_{p})(-2i  +\Delta_{c}-\Delta_{p})](1+\Delta_{c}^{2}+2\Omega_{c}^{2})-iP^{2}\{\Delta_{p}(5-3i\Delta_{p})\nonumber\\&&+\Delta_{c}^{2}(3i+\Delta_{p})+\Omega_{c}^{2}(2i-3\Delta_{p})+\Delta_{c}[9-\Delta_{p}(8i +3\Delta_{p})+\Omega_{c}^{2}] \}, \nonumber\\
&B_{4}=&8iP^{4}+P^{2}\{-2i[8+\Delta_{c}(-4i+\Delta_{c})-4i\Delta_{p}-6\Delta_{c}\Delta_{p}+\Delta_{p}^{2}]+\Omega_{c}^{2}(6i+\Delta_{c}-5\Delta_{p}) \} \nonumber\\
        &&+ [\Omega_{c}^{2}+(i+\Delta_{p})(-2i+\Delta_{c}-\Delta_{p})][2(1-i\Delta_{c})(2i+\Delta_{c}-\Delta_{p})+\Omega_{c}^{2}(3i+\Delta_{c}-2\Delta_{p})],\\ \nonumber
&B_{5}=&P^{4}(4i+\Delta_{c}-3\Delta_{p})+[\Omega_{c}^{2}+(i+\Delta_{p})(-2i+\Delta_{c}-\Delta_{p})][2(i+\Delta_{c})(3+i\Delta_{p})+\Omega_{c}^{2}(7i+\Delta_{c}-2\Delta_{p})] \nonumber\\
       &&+P^{2}\{\Delta_{c}^{2}(i+\Delta_{p})+\Delta_{p}(-3-5i\Delta_{p}- 8\Omega_{c}^{2})+8i(-2+\Omega_{c}^{2})+\Delta_{c}[1+(4i  -3\Delta_{p})\Delta_{p}+2\Omega_{c}^{2}]\}, \nonumber\\
&B_{6}=&P^{2}(6i+\Delta_{c}-5\Delta_{p})+(5i+\Delta_{c}-2\Delta_{p})[(-2i+\Delta_{c}-\Delta_{p})(i+\Delta_{p})+\Omega_{c}^{2}],  \nonumber\\
&B_{7}=&-4iP^{5}\Omega_{c}+iP^{3}[8+\Delta_{c}^{2}+\Delta_{p}(-4i+\Delta_{p})-2\Delta_{c}(2i+3\Delta_{p})]\Omega_{c}+p\Omega_{c}[(2+i\Delta_{c}-i\Delta_{p}+i\Omega_{c}^{2})(i+ \nonumber\\
        &&\Delta_{p})][(i+\Delta_{c})(2i+\Delta_{c}-\Delta_{p})+\Omega_{c}^{2}]+(1+\Delta_{c}^{2}+2\Omega_{c}^{2})+\{[4+(\Delta_{c}-\Delta_{p})^{2}](i+\Delta_{p})+(2i+\Delta_{c} \nonumber\\
        &&-\Delta_{p})\Omega_{c}^{2}\}\Omega_{p}+P^{2}\{(\Delta_{c}-i)[8+\Delta_{c})^{2}+\Delta_{p}(\Delta_{p}-4i)-2\Delta_{c}(2i+3\Delta_{p})]-[12i+5\Delta_{c}+(3+2i\Delta_{p}) \nonumber\\
        &&\Delta_{p}]\Omega_{c}^{2}- i\Omega_{c}^{4}\}\Omega_{p}+2P^{4}[-2\Delta_{c}+i(2+\Omega_{c}^{2})]\Omega_{p}, \nonumber\\
&B_{8}=&-2\Delta_{c}^{3}\Delta_{p}+6(2+\Delta_{p})^{2}-2\Delta_{c}\Delta_{p}(6+\Delta_{p})^{2})+\Delta_{c}^{2}(6+8\Delta_{p})^{2}+2[4-(\Delta_{c}-4\Delta_{p})(Delta_{c}+\Delta_{p})]\Omega_{c}^{2}+\Omega_{c}^{4}, \nonumber\\
&B_{9}=&[4+(\Delta_{c}-\Delta_{p})^{2}](1+\Delta_{p})^{2})+2[2+(\Delta_{c}-\Delta_{p})\Delta_{p}]\Omega_{c}^{2}+\Omega_{c}^{4},  \nonumber\\
&B_{10}=&12+\Delta_{c}^{2}-10\Delta_{c}\Delta_{p}+\Delta_{p}^{2}-4\Omega_{c}^{2}. \nonumber
\end{eqnarray}\end{widetext}

\end{document}